\documentclass[conference]{IEEEtran}
\usepackage{epsfig}
\usepackage{amssymb}
\usepackage{amsmath}
\usepackage{latexsym,graphicx}
\usepackage{longtable}
\usepackage{algpseudocode} 
\usepackage{algorithm}

\usepackage{array}
\usepackage{mathrsfs}
\usepackage{color}
\usepackage{amsmath,graphicx}
\usepackage{epsfig}
\usepackage{amssymb}
\usepackage{setspace}
\usepackage{cite}
\usepackage{hyperref}
\usepackage{amssymb}
\usepackage{caption}
\usepackage{subcaption}
\def\BibTeX{{\rm B\kern-.05em{\sc i\kern-.025em b}\kern-.08em
    T\kern-.1667em\lower.7ex\hbox{E}\kern-.125emX}}
 \makeatletter
 \let\NAT@parse\undefined
 \makeatother

\begin{document}

\title{Energy-Efficient Data Collection and Wireless Power Transfer Using A MIMO Full-Duplex UAV}
%
%
\author{$\text{Jiancao Hou}$, $\text{Zhaohui Yang}$ and $\text{Mohammad Shikh-Bahaei}$
\\{\small Centre for Telecommunications Research, King's College London, London, UK}
\\{\small e-mail:\{jiancao.hou, yang.zhaohui, m.sbahaei\}@kcl.ac.uk}


}
\markboth{} {Shell \MakeLowercase{\textit{et al.}}: Bare Demo of
IEEEtran.cls for Journals}\maketitle

\begin{abstract}
In this paper, we propose a novel energy-efficient data collection and wireless power transfer (WPT) framework for internet of things (IoT) applications, via a multiple-input multiple-output (MIMO) full-duplex (FD) unmanned aerial vehicle (UAV). To exploit the benefits of UAV-enabled WPT and MIMO FD communications, we allow the MIMO FD UAV charge low-power IoT devices while at the same time collect data from them. With the aim of saving the total energy consumed at the UAV, we formulate an energy minimization problem by taking the FD hardware impairments, the number of uploaded data bits, and the energy harvesting causality into account. Due to the non-convexity of the problem in terms of UAV trajectory and transmit beamforming for WPT, tracking the global optimality is quite challenge. Alternatively, we find a local optimal point by implementing the proposed iterative search algorithm combining with successive convex approximation techniques. Numerical results show that the proposed approach can lead to superior performance compared with other benchmark schemes with low computational complexity and fast convergence.    

\end{abstract}


\IEEEpeerreviewmaketitle
 

\section{Introduction}
Wireless communications are rapidly evolving into a massive internet of things (IoT) environment, where a large amount of sensors are expected to communicate relying on low latency and high energy and/or spectral efficiencies \cite{Mozaffari2018,Bobarshad2009,Orikumhi2017}. With such kind of consideration, many IoT applications, e.g., data collection from insect traps deployed in farmland, periodically deliver the obtained data of IoT devices to a central processing unit. However, due to their energy constraints, the IoT devices are highly battery limited and are typically unable to transmit over a long distance. The use of low-altitude unmanned aerial vehicles (UAVs) can provide a cost-effective and energy-efficient solution to collect data from the ground IoT devices that are spread over a geographical area \cite{Zeng2017,Mozaffari2017}. In this case, the UAV(s) can dynamically move towards IoT devices and collect data from them. Here, the UAV(s) play the role of moving base station(s), and to effectively design their trajectories and reduce the energy consumptions are quite challenge tasks \cite{Yang2018}. 

In-band full-duplex (IBFD) as one of promising transmission techniques has recently re-emerged to improve system spectral efficiency \cite{Sabharwal2014,Kim2015,Towhidlou2016,Towhidlou2018,Shadmand2009}. Unlike the conventional half-duplex radio transceiver, IBFD is able to transmit and receive simultaneously over the same frequency band. Such technique can be implemented at the UAV(s) to collect data and charge low-power IoT devices at the same time. In theory, this approach can halve communication latency and double spectral efficiency if the self-interference (SI) is well eliminated \cite{Naslcheraghi2017,Febrianto2017}. Many research work has been conducted to develop advanced techniques to handle the SI effect, including spatial domain SI cancellation \cite{Choi2010,Everett2014}. However, in practice to completely remove the SI effect is challenging due to the inaccuracy of the hardware components, e.g., limited resolutions of analog-to-digital/digital-to-analog converters, power amplifier and oscillator phase noise \cite{Sabharwal2014,Taghizadeh2018}. As a result, it is essential to take these hardware impairments into account to design an efficient FD communication system under realistic situations. In the work on 4G support for low-altitude UAV(s), another big problem is due to multi-user interference caused by simultaneously supported multiple IoT devices \cite{ATIS2018}. In this case, multiple-input multiple-output (MIMO) beamforming can be implemented at UAV(s) to cancel the co-channel interference and also provide efficient wireless power transfer (WPT) \cite{Wu2018}. 

Motivated by the above discussion, in this paper, we propose an energy-efficient data collection and WPT framework for simultaneously supporting multiple IoT devices by using a MIMO FD UAV. Specifically, assume that the transmit power of the supported IoT devices are purely from WPT of the UAV, we aim to minimize the energy consumption at the UAV by taking the FD hardware impairments, the number of uploaded data bits, and the energy harvesting causality into account. In the literature, the energy-efficient UAV communications with trajectory optimization has been investigated in \cite{Zeng2017,Mozaffari2017}. In addition, the joint computation offloading and trajectory optimization has also been studied in \cite{Cao2018,Jeong2018,Zhou2018}, where WPT technique is considered in \cite{Zhou2018}. To the best of our knowledge, there is no much work to study data collection and WPT via a MIMO FD UAV in consideration of hardware impairments, and due to non-convexity of the formulated problem with respect to UAV trajectory and the covariance matrix of its transmitted energy signal, tracking the global optimality is quite challenge. Thus, we propose an iterative search combining with successive convex approximation (SCA) method to find a local optimal solution. Numerical results show that the proposed algorithm outperforms other benchmark schemes in terms energy consumption with low computational complexity and fast convergence.



\section{Hardware Impairments Aware System Modelling}
Consider a UAV enabled data collection and WPT framework, 
where a rotary wing UAV with FD capability and $M$ number of antennas simultaneously serve $K$ ($\leq M$) single-antenna based FD IoT devices (e.g., insect traps in farmland). The IoT devices suppose to upload their obtained data, i.e., $R_k,_{k=1,\ldots K}$, bits to the UAV by using their harvested energy from the UAV. The UAV uses all its $M$ antennas to form an energy beam to charge $K$ IoT devices, and at the same time to collect data from them. We assume the UAV perfectly know the location of each IoT device for designing trajectory, and every node has channel knowledge of its related links including the SI channel if any.

Without loss of generality, a three-dimensional (3D) Euclidean space is adopted to present the locations of $K$ IoT devices and the UAV. Assume that all IoT devices are fixed at ground, and their locations are denoted as $\mathbf{q}_{k}\triangleq[x_k,y_k],_{k=1,\ldots,K}$, where $x_k$ and $y_k$ are the horizontal plane coordinate of the $k^{\mathrm{th}}$ IoT device. Moreover, the UAV flies at a fixed altitude $L$ from the ground and collect data with a finite time $T$. For ease of exposition, we divide the time $T$ into $N$ equal-time slots, and at each time slot, the UAV hovers at a fix position for data collection and WPT. We assume the channels between the UAV and the IoT devices during a divided time slot are unchanged. Let's denote $\mathbf{q}_{\mathrm{u}}[n]\triangleq[x_{\mathrm{u}}[n],y_{\mathrm{u}}[n]],_{n=1,\ldots N}$ as the horizontal plane coordinate of the UAV. Thus, the channels between the UAV and the $k^{\mathrm{th}}$ IoT device can be formulated as the free-space path loss
\begin{equation}\label{eq01}
r_{k}[n]=\frac{1}{L^2+\|\mathbf{q}_{\mathrm{u}}[n]-\mathbf{q}_{k}\|^{2}},~\forall k,n,
\end{equation}
multiplied by the small scale fading effect, i.e., $\mathbf{h}_{k}[n]\in\mathcal{C}^{M\times1}$, which is subject to Rician fading with elements distributed independently as $\mathcal{CN}(\mu,\nu^2)$.\footnote{The Rician channel assumption is due to the line-of-sight (LOS) component between the UAV and $K$ IoT devices.  } In addition, due to FD capability, the received signals of the UAV are interfered by the energy bearing signal transmitted by itself, and the loop-back SI channel, i.e., $\mathbf{H}_{\mathrm{u}}[n]\in\mathcal{C}^{M\times M}$ can also be modelled as Rician distribution with $\mu_\mathrm{u}$ mean and $\nu^2_{\mathrm{u}}$ variance.     

To investigate energy consumptions of the MIMO FD UAV, as aforementioned, the hardware impairments should be taken into account \cite{Sabharwal2014,Taghizadeh2018}. Let's denote $\mathbf{e}_{\mathrm{in}}[n]\in\mathcal{C}^{M\times1}$ as the distortion at the receiver side, which represents combined effect of UAV receiver radio frequency (RF) chain impairments; $\mathbf{e}_{\mathrm{out}}[n]\in\mathcal{C}^{M\times1}$ as the distortion at the transmitter side, which represents combined effect of UAV transmitter RF chain impairments. We assume $\mathbf{x}[n]\in\mathcal{C}^{M\times1}$ as the energy-bearing signal sent by the UAV for charging the $K$ IoT devices in the $n^{\mathrm{th}}$ time slot, and $\mathbf{X}[n]\triangleq\mathbb{E}\{\mathbf{x}[n]\mathbf{x}^{H}[n]\}$ denotes its covariance matrix,\footnote{In general, the UAV can generate multiple energy beams for the WPT, thus, $\mathbf{X}[n],_{\forall n}$ can be of any rank \cite{Wu2018,Xing2017}.} where $\mathbb{E}\{\cdot\}$ denotes the expectation operation and $(\cdot)^{H}$ denotes the conjugate transpose. Thus, the received signal at the UAV can be expressed as
\begin{IEEEeqnarray}{ll}\label{eq02}
\mathbf{y}[n]&=\sum^{K}_{k=1}\sqrt{P_{k}[n]r_k[n]}\mathbf{h}_k[n]s_k+\mathbf{H}_{\mathrm{u}}[n](\mathbf{x}[n]+\mathbf{e}_{\mathrm{out}}[n])\nonumber\\
&~~~+\mathbf{v}[n]+\mathbf{e}_{\mathrm{in}}[n],~\forall n,
\end{IEEEeqnarray}
where $P_{k}[n]$ is the transmit power at the $k^{\mathrm{th}}$ IoT device in the $n^{\mathrm{th}}$ time slot; $s_{k}$ is the transmitted data symbol from the $k^{\mathrm{th}}$ IoT device and $\mathbb{E}\{|s_k|^2\}=1$; $\mathbf{v}[n]\in\mathcal{C}^{M\times1}$ is the additive white Gaussian noise (AWGN) at the UAV in the $n^{\mathrm{th}}$ time slot with zero mean and $\sigma^{2}_{0}$ variance. Furthermore, according to \cite{Taghizadeh2018}, the transceiver distortions, i.e., $\mathbf{e}_{\mathrm{in}}[n]$ and $\mathbf{e}_{\mathrm{out}}[n]$, can be expressed as  
\begin{IEEEeqnarray}{ll}\label{eq03}
\mathbf{e}_{\mathrm{in}}[n]\sim\mathcal{CN}\left(\mathbf{0},\beta\mathrm{diag}\left(\mathbb{E}\{\mathbf{u}[n]\mathbf{u}^{H}[n]\}\right)\right),~\forall n,
\end{IEEEeqnarray}
and
\begin{IEEEeqnarray}{ll}\label{eq04}
\mathbf{e}_{\mathrm{out}}[n]\sim\mathcal{CN}\left(\mathbf{0},\kappa\mathrm{diag}\left(\mathbf{X}[n]\}\right)\right),~\forall n,
\end{IEEEeqnarray}
where $\mathbf{u}[n]$ denotes undistorted receive signal at the UAV, which is composed of the first three terms on the right hand side of \eqref{eq02}; $\beta,\kappa\in\mathcal{R}^{+}$ are the receive and transmit distortion coefficients, respectively, which reflect the quality of the SI cancellation in propagation/analog domains. In \eqref{eq03}, the covariance matrix of $\mathbf{u}[n]$ is given by
\begin{IEEEeqnarray}{ll}\label{eq05}
\mathbb{E}\{\mathbf{u}[n]\mathbf{u}^{H}[n]\}&=\sum^{K}_{k=1}P_{k}[n]r_k[n]\mathbf{h}_k[n]\mathbf{h}^{H}_{k}[n]+\sigma^{2}_{0}\mathbf{I}\nonumber\\
&~~~+\mathbf{H}_{\mathrm{u}}[n]\mathbb{E}\{\mathbf{w}[n]\mathbf{w}^{H}[n]\}\mathbf{H}^{H}_{\mathrm{u}}[n],~\forall n.
\end{IEEEeqnarray}
where $\mathbf{w}[n]\triangleq\mathbf{x}[n]+\mathbf{e}_{\mathrm{out}}[n]$, and $\mathbf{I}$ denotes the identity matrix. 
In practice, the loop-back SI signal is much stronger than the received signals from IoT devices. Hence, we approximate 
\begin{IEEEeqnarray}{ll}\label{eq06}
\mathbb{E}\{\mathbf{u}[n]\mathbf{u}^{H}[n]\}&\approx\mathbf{H}_{\mathrm{u}}[n]\mathbb{E}\{\mathbf{w}[n]\mathbf{w}^{H}[n]\}\mathbf{H}^{H}_{\mathrm{u}}[n]\nonumber\\
&\overset{(a)}{=}\mathbf{H}_{\mathrm{u}}[n](\mathbf{X}[n]+\kappa\mathrm{diag}(\mathbf{X}[n]))\mathbf{H}^{H}_{\mathrm{u}}[n],\forall n,
\end{IEEEeqnarray}
where step-(a) in \eqref{eq06} is obtained by recalling $\mathbf{X}[n]$ and \eqref{eq04}. 

To completely remove the multi-user interference and simplify the detection process, the receiver side of the UAV will implement a linear zero-forcing (ZF) technique, and the post-processing matrix is formulated as \cite{Goldsmith2005}
\begin{IEEEeqnarray}{ll}\label{eq07}
\mathbf{Z}[n]=(\mathbf{H}^{H}[n]\mathbf{H}[n])^{-1}\mathbf{H}^{H}[n],~\forall n.
\end{IEEEeqnarray}
In \eqref{eq07}, $\mathbf{H}[n]\triangleq[\mathbf{h}_{1}[n],\ldots,\mathbf{h}_{K}[n]]$ is the combination channel matrix between the $K$ IoT devices and the UAV in the $n^{\mathrm{th}}$ time slot. Here, the $k^{\mathrm{th}}$ row of $\mathbf{Z}[n]$, i.e., $\mathbf{z}_{k}[n]\in\mathcal{C}^{1\times M}$, is the receive beamforming vector for the $k^{\mathrm{th}}$ IoT device in the $n^{\mathrm{th}}$ time slot. In addition, the SI cancellation at the UAV in digital domain can be done by subtracting $\mathbf{H}_{\mathrm{u}}[n]\mathbf{x}[n]$ from \eqref{eq02}. Then, the received signal at
 the UAV for the $k^{\mathrm{th}}$ IoT device in the $n^{\mathrm{th}}$ time slot can be expressed as
\begin{IEEEeqnarray}{ll}\label{eq08}
\mathbf{z}_{k}[n]\mathbf{y}_{k}[n]&=\sqrt{P_{k}[n]r_k[n]}\mathbf{z}_{k}[n]\mathbf{h}_k[n]s_k+\mathbf{z}_{k}[n]\mathbf{H}_{\mathrm{u}}[n]\mathbf{e}_{\mathrm{out}}[n]\nonumber\\
&~~~+\mathbf{z}_{k}[n]\mathbf{v}[n]+\mathbf{z}_{k}[n]\mathbf{e}_{\mathrm{in}}[n],~\forall k, n,
\end{IEEEeqnarray}
and the signal-to-interference plus noise ratio (SINR) for the $k^{\mathrm{th}}$ IoT device in the $n^{\mathrm{th}}$ time slot can be formulated as
\begin{IEEEeqnarray}{ll}\label{eq09}
\mathrm{SINR}_{k}[n]=~~~~~~~~~~~~~~~~~~~~~~~~~~~~~~~~~~~~~~~~~~~~~~~~~~~~~~~~\nonumber\\
\frac{P_{k}[n]r_k[n]|\mathbf{z}_{k}[n]\mathbf{h}_k[n]|^2}{\mathbf{z}_{k}[n]\left(\mathbf{H}_{\mathrm{u}}[n]\mathbf{E}_{\mathrm{out}}[n]\mathbf{H}^{H}_{\mathrm{u}}[n]+\sigma^{2}_{0}\mathbf{I}+\mathbf{E}_{\mathrm{in}}[n]\right)\mathbf{z}^{H}_{k}[n]},~\forall k,n,\nonumber\\
\end{IEEEeqnarray}
where 
\begin{eqnarray}
\mathbf{E}_{\mathrm{out}}[n]\triangleq\mathbb{E}\{\mathbf{e}_{\mathrm{out}}[n]\mathbf{e}^{H}_{\mathrm{out}}[n]\}=\kappa\mathrm{diag}(\mathbf{X}[n]),\nonumber
\end{eqnarray}
and
\begin{IEEEeqnarray}{ll}
\mathbf{E}_{\mathrm{in}}[n]&\triangleq\mathbb{E}\{\mathbf{e}_{\mathrm{in}}[n]\mathbf{e}^{H}_{\mathrm{in}}[n]\}\nonumber\\
&\approx\beta\mathrm{diag}\left(\mathbf{H}_{\mathrm{u}}[n](\mathbf{X}[n]+\kappa\mathrm{diag}(\mathbf{X}[n]))\mathbf{H}^{H}_{\mathrm{u}}[n]\right).\nonumber
\end{IEEEeqnarray}

From the IoT devices point of view, we assume that the system works in time division duplex mode. In this case, the received signal at the $k^{\mathrm{th}}$ IoT device in the $n^{\mathrm{th}}$ time slot can be expressed as
\begin{IEEEeqnarray}{ll}\label{eq10}
y_{k}[n]=\sqrt{r_{k}[n]}\mathbf{h}^{T}_{k}[n](\mathbf{x}[n]+\mathbf{e}_{\mathrm{out}}[n]),~\forall k,n,
\end{IEEEeqnarray}
where $(\cdot)^{T}$ denotes the transpose. Then, its harvest power can be expressed as
\begin{IEEEeqnarray}{ll}\label{eq11}
P_{\mathrm{hp},k}[n]=\eta r_{k}[n]\mathbf{h}^{T}_{k}[n](\mathbf{X}[n]+\kappa\mathrm{diag}(\mathbf{X}[n]))\mathbf{h}_{k}[n],~\forall k,n,\nonumber\\
\end{IEEEeqnarray}
where $\eta$ denotes the harvest power conversion efficiency. It is worth noting the received SI, hardware impairments, and AWGN at IoT devices are ignored in \eqref{eq10} due to their limited contributions on energy harvesting.



\section{Problem Formulation}
The energy consumption at the UAV mainly comes from two parts: the propulsion energy and the WPT. For the first part, denote $t$ as the UAV moving time from one location to another. Then, similar to the work in \cite{Jeong2018,Zhou2018}, the UAV propulsion energy in the $n^{\mathrm{th}}$ time slot can be expressed as 
\begin{IEEEeqnarray}{ll}\label{eq12}
E_{\mathrm{p}}[n]=\tau|v_{\mathrm{u}}[n]|^{2},~\forall n,
\end{IEEEeqnarray}
where $v_{\mathrm{u}}[n]\triangleq{\|\mathbf{q}_{\mathrm{u}}[n+1]-\mathbf{q}_{\mathrm{u}}[n]\|}/{t}$ denotes velocity of the UAV and $\tau$ denotes propulsion energy efficiency. The second part is the transmit power at the UAV for WPT, which is
\begin{IEEEeqnarray}{ll}\label{eq13}
P_{\mathrm{wpt}}[n]=\mathrm{trace}(\mathbf{X}[n]+\kappa\mathrm{diag}(\mathbf{X}[n])),~\forall n,
\end{IEEEeqnarray}
where $\mathrm{trace}(\cdot)$ denotes the trace operation of a matrix. Thus, the UAV energy minimization problem can be formulated as
\begin{IEEEeqnarray}{ll}\label{eq14}
\underset{P_{k}[n],\mathbf{X}[n],\mathbf{q}_{\mathrm{u}}[n]}{\mathrm{minimize}}&~~E_{\mathrm{p}}[0]+\sum^{N}_{n=1}(E_{\mathrm{p}}[n]+\frac{T}{N}P_{\mathrm{wpt}}[n])\nonumber\\\label{eq14a}
~~~\mathrm{subject~to}&~~\sum^{N}_{n=1}\frac{T}{N}B\log_{2}(1+\mathrm{SINR}_{k}[n])\geq R_{k},~\forall k,
\nonumber\\\IEEEyesnumber
\IEEEyessubnumber\\\label{eq14b}
&~~\sum^{n}_{i=1}\frac{T}{N}P_{k}[i]\leq\sum^{n}_{i=1}\frac{T}{N}P_{\mathrm{hp},k}[i],~\forall k,n,
\IEEEyesnumber
\IEEEyessubnumber\\\label{eq14c}
&~~\mathbf{q}_{\mathrm{u}}[0]=\mathbf{q}_{\mathrm{ui}},~\mathbf{q}_{\mathrm{u}}[N+1]=\mathbf{q}_{\mathrm{uf}},
\IEEEyesnumber
\IEEEyessubnumber\\\label{eq14d}
&~~v_{\mathrm{u}}[n]\leq V_{\mathrm{max}},~P_{\mathrm{wpt}}[n]\leq P_\mathrm{max},~\forall n,
\IEEEyesnumber
\IEEEyessubnumber
\end{IEEEeqnarray}
where $B$ denotes the available bandwidth for transmissions; $\mathbf{q}_{\mathrm{ui}}$ and $\mathbf{q}_{\mathrm{uf}}$ are initial and final positions of the UAV, respectively; $V_{\mathrm{max}}$ and $P_{\mathrm{max}}$ denote the maximum speed and the maximum transmit power of the UAV, respectively.

In problem \eqref{eq14}, (14a) illustrates that the total uploaded bits for the $k^{\mathrm{th}}$ IoT device should be no less than $R_{k}$. As mentioned before, the transmit power of the IoT devices are solely supplied by its harvest power from the UAV, thus, (14b) illustrates that the total transmit energy at each IoT device until the $n^{\mathrm{th}}$ time slot is no larger than its harvested energy. In addition, (14c) sets up initial and final positions of the UAV, and (14d) limits speed and transmit power of the UAV. 
   


\section{The Proposed Joint Trajectory and Transmit Beamforming Design}
In this section, we aim to solve problem \eqref{eq14}. Specifically, problem \eqref{eq14} is non-convex in terms of both $\mathbf{X}[n]$ and $\mathbf{q}_{\mathrm{u}}[n]$. In addition, due to the coupling among $P_{k}[n]$, $\mathbf{X}[n]$ and $\mathbf{q}_{\mathrm{u}}[n]$, finding the globally optimal point is quite challenge. Alternatively, we propose an iterative search method to find a locally optimal solution, where the first loop is to find the optimal $P_{k}[n]$ and $\mathbf{X}[n]$ by fixing $\mathbf{q}_{\mathrm{u}}[n]$, and the second loop is to find the optimal $\mathbf{q}_{\mathrm{u}}[n]$ by fixing $P_{k}[n]$ and $\mathbf{X}[n]$.
  
\subsection{The First Loop Optimization}
By fixing $\mathbf{q}_{\mathrm{u}}[n],_{\forall n}$ and introducing auxiliary variables $t_{k}[n],_{\forall k,n}\geq0$, we can reformulate problem \eqref{eq14} as
\begin{IEEEeqnarray}{ll}\label{eq15}
\underset{P_{k}[n],\mathbf{X}[n],t_{k}[n]}{\mathrm{minimize}}&~~\sum^{N}_{n=1}P_{\mathrm{wpt}}[n]\nonumber\\
~~~\mathrm{subject~to}&~~\sum^{N}_{n=1}\frac{T}{N}B\log_{2}(1+t_{k}[n])\geq R_{k},~\forall k,
\IEEEyesnumber
\IEEEyessubnumber\\
&~~\mathrm{SINR}_{k}[n]\geq t_{k}[n],~\forall k,n,
\IEEEyesnumber
\IEEEyessubnumber\\
&~~P_{\mathrm{wpt}}[n]\leq P_\mathrm{max},~\forall n,
\IEEEyesnumber
\IEEEyessubnumber\\
&~~(14\mathrm{b}).\nonumber
\end{IEEEeqnarray}
Problem \eqref{eq15} is non-convex due to coupling $\mathbf{X}[n]$ and $t_{k}[n]$ in constraint (15b) for all $k$ and $n$. To facilitate solving \eqref{eq15}, we decouple the numerator and the denominator of left hand side of (15b) by introducing additional auxiliary variables $e_{k}[n],_{\forall k,n}\geq0$, as
\begin{IEEEeqnarray}{ll}\label{eq16}
P_{k}[n]r_k[n]|\mathbf{z}_{k}[n]\mathbf{h}_k[n]|^2\geq t_{k}[n]e_{k}[n], ~\forall k,n,
\end{IEEEeqnarray}
\begin{IEEEeqnarray}{ll}\label{eq17}
\mathbf{z}_{k}[n]\left(\mathbf{H}_{\mathrm{u}}[n]\mathbf{E}_{\mathrm{out}}[n]\mathbf{H}^{H}_{\mathrm{u}}[n]+\sigma^{2}_{0}\mathbf{I}+\mathbf{E}_{\mathrm{in}}[n]\right)\mathbf{z}^{H}_{k}[n]\leq e_{k}[n],\nonumber\\~~~~~~~~~~~~~~~~~~~~~~~~~~~~~~~~~~~~~~~~~~~~~~~~~~~~~~~~~~~\forall k,n.
\end{IEEEeqnarray}
Due to $t_{k}[n]e_{k}[n]$ in \eqref{eq16}, the problem is still non-convex. Thus, the first-order Taylor expansion of $t_{k}[n]e_{k}[n]$ gives
\begin{IEEEeqnarray}{ll}
t_{k}[n]e_{k}[n]\nonumber\\
\leq\overline{t}_{k}[n]\overline{e}_{k}[n]+\overline{e}_{k}[n](t_{k}[n]-\overline{t}_{k}[n])+\overline{t}_{k}[n](e_{k}[n]-\overline{e}_{k}[n]),\nonumber
\end{IEEEeqnarray}
where the expansion is near the initial point pair $(\overline{t}_{k}[n],\overline{e}_{k}[n])$, and equality holds if and only if $t_{k}[n]=\overline{t}_{k}[n]$ and $e_{k}[n]=\overline{e}_{k}[n]$. Then, \eqref{eq16} can be approximated as 
\begin{IEEEeqnarray}{ll}\label{eq18}
P_{k}[n]r_k[n]|\mathbf{z}_{k}[n]\mathbf{h}_k[n]|^2\nonumber\\
\geq\overline{t}_{k}[n]\overline{e}_{k}[n]+\overline{e}_{k}[n](t_{k}[n]-\overline{t}_{k}[n])+\overline{t}_{k}[n](e_{k}[n]-\overline{e}_{k}[n]),\nonumber\\~~~~~~~~~~~~~~~~~~~~~~~~~~~~~~~~~~~~~~~~~~~~~~~~~~~~~~~~~~~\forall k,n.
\end{IEEEeqnarray}

With the above discussed process, the non-convex problem \eqref{eq15} can be converted to the convex problem as
\begin{IEEEeqnarray}{ll}\label{eq19}
\underset{P_{k}[n],\mathbf{X}[n],t_{k}[n],e_{k}[n]}{\mathrm{minimize}}&~~\sum^{N}_{n=1}P_{\mathrm{wpt}}[n]\\
~~~~~~\mathrm{subject~to}&~~(15\mathrm{a}), (17), (18), (15\mathrm{c}), (14\mathrm{b}), \nonumber
\end{IEEEeqnarray}
where it can be readily solved by using the optimization toolboxes, e.g., cvx in \cite{Grant2015}. Then, by fixing $\mathbf{q}_{\mathrm{u}}[n],_{\forall n}$ and providing feasible $(\overline{t}_{k}[n],\overline{e}_{k}[n])$ pair for all $k$ and $n$, the optimal $P^{*}_{k}[n]$ and $\mathbf{X}^{*}[n]$ for all $k$ and $n$ can be found by solving problem \eqref{eq19} via standard SCA techniques \cite{Facchinei2015}.

\subsection{The Second Loop Optimization}
Given $P_{k}[n]$ and $\mathbf{X}[n]$, let's first treat $r_{k}[n],_{\forall k,n,}$ as auxiliary variables and replace `$=$' with `$\leq$' from formula \eqref{eq01}. Then, the problem \eqref{eq14} can be reformulated as
\begin{IEEEeqnarray}{ll}\label{eq20}
\underset{\mathbf{q}_{\mathrm{u}}[n],r_{k}[n]}{\mathrm{minimize}}&~~\sum^{N}_{n=0}E_{\mathrm{p}}[n]\nonumber\\
~\mathrm{subject~to}&~~\frac{1}{L^2+\|\mathbf{q}_{\mathrm{u}}[n]-\mathbf{q}_{k}\|^{2}}\geq r_{k}[n],~\forall k,n,
\IEEEyesnumber
\IEEEyessubnumber\\
&~~v_{\mathrm{u}}[n]\leq V_{\mathrm{max}},~\forall n,
\IEEEyesnumber
\IEEEyessubnumber\\
&~~(14\mathrm{a}), (14\mathrm{b}), (14\mathrm{c}).\nonumber
\end{IEEEeqnarray}
Problem \eqref{eq20} is non-convex due to coupling $\mathbf{q}_{\mathrm{u}}[n]$ and $r_{k}[n]$ in constraint (20a) for all $k$ and $n$. On the other hand, due to the introduced auxiliary variables, i.e., $r_{k}[n],_{\forall k,n}$, both (14a) and (14b) become convex constraints. Similar as \eqref{eq16} and \eqref{eq17}, by introducing additional auxiliary variables $f_{k}[n]_{\forall k,n}$, we can convert (20a) into 
\begin{IEEEeqnarray}{ll}\label{eq21}
r_{k}[n]f_{k}[n]\leq1,~\forall k,n,
\end{IEEEeqnarray}
\begin{IEEEeqnarray}{ll}\label{eq22}
L^2+\|\mathbf{q}_{\mathrm{u}}[n]-\mathbf{q}_{k}\|^{2}\leq f_{k}[n],~\forall k,n.
\end{IEEEeqnarray}
Then, the first-order Taylor expansion of $r_{k}[n]f_{k}[n]$ gives
\begin{IEEEeqnarray}{ll}
r_{k}[n]f_{k}[n]\nonumber\\
\leq\overline{r}_{k}[n]\overline{f}_{k}[n]+\overline{f}_{k}[n](r_{k}[n]-\overline{r}_{k}[n])+\overline{r}_{k}[n](f_{k}[n]-\overline{f}_{k}[n]),\nonumber
\end{IEEEeqnarray}
where the expansion is near the initial point pair $(\overline{r}_{k}[n],\overline{f}_{k}[n])$, and equality holds if and only if $r_{k}[n]=\overline{r}_{k}[n]$ and $f_{k}[n]=\overline{f}_{k}[n]$. Then, \eqref{eq21} can be approximated as 
\begin{IEEEeqnarray}{ll}\label{eq23}
\overline{r}_{k}[n]\overline{f}_{k}[n]+\overline{f}_{k}[n](r_{k}[n]-\overline{r}_{k}[n])+\overline{r}_{k}[n](f_{k}[n]-\overline{f}_{k}[n])\nonumber\\~~~~~~~~~~~~~~~~~~~~~~~~~~~~~~~~~~~~~~~~~~~~~~~~~~\leq1,~\forall k,n.
\end{IEEEeqnarray}

Following the above discussion by taking steps \eqref{eq21} to \eqref{eq23} into account, the non-convex problem \eqref{eq20} can be converted into the convex problem as
\begin{IEEEeqnarray}{ll}\label{eq24}
\underset{\mathbf{q}_{\mathrm{u}}[n],r_{k}[n],f_{k}[n]}{\mathrm{minimize}}&~~\sum^{N}_{n=1}E_{\mathrm{p}}[n]\\
~~~\mathrm{subject~to}&~~(22), (23), (20\mathrm{b}), (14\mathrm{a}), (14\mathrm{b}), (14\mathrm{c}). \nonumber
\end{IEEEeqnarray}
Then, the method used to solve problem \eqref{eq19} can be used to solve problem \eqref{eq24}. 

\subsection{The Proposed Iterative Search Algorithm}
Before we introduce the proposed iterative search algorithm, it is worth to highlight that proper chosen of initial ($\overline{t}_{k}[n]$,$\overline{e}_{k}[n]$) and ($\overline{r}_{k}[n]$,$\overline{f}_{k}[n]$) for all $k$ and $n$ are necessary to efficiently solve problems \eqref{eq19} and \eqref{eq24}, respectively. For finding initial ($\overline{t}_{k}[n]$,$\overline{e}_{k}[n]$), we first set $t_{k}[n]=t,\forall k,n$, and obtain $t$ by solving $\sum^{N}_{n=1}\frac{T}{N}B\log_{2}(1+t)=R_{k},\forall k$. Here, the obtained $t$ will be treated as the initial $\overline{t}_{k}[n],_{\forall k, n}$. Then, given the initial $\overline{t}_{k}[n]_{\forall k, n}$ and $\mathbf{q}_{\mathrm{u}}[n]_{\forall n}$, the initial $\overline{e}_{k}[n]_{\forall k,n,}$ can be found by solving the following problem
\begin{IEEEeqnarray}{ll}\label{eq24}
\underset{P_{k}[n],\mathbf{X}[n],e_{k}[n]}{\mathrm{minimize}}&~~0\\
~~~\mathrm{subject~to}&~~(16), (17), (15\mathrm{c}), (14\mathrm{b}). \nonumber
\end{IEEEeqnarray}
For the second loop optimization process, the initial pair ($\overline{r}_{k}[n]$,$\overline{f}_{k}[n]$) can be formulated by letting $\overline{r}_{k}[n]=\frac{1}{L^2+\|\mathbf{q}_{\mathrm{u}}[n]-\mathbf{q}_{k}\|^{2}}$ and $\overline{f}_{k}[n]=L^2+\|\mathbf{q}_{\mathrm{u}}[n]-\mathbf{q}_{k}\|^{2}$ for all $k$ and $n$. Consequently, the proposed iterative search algorithm can be formulated as follows.
\begin{algorithm}
\caption{The Proposed Iterative Search Algorithm}
\begin{algorithmic}[1]
\State\textbf{Initialize:} $\mathbf{q}^{(0)}_{\mathrm{u}}[n]$, $\overline{t}^{(0)}_{k}[n]$, $\overline{e}^{(0)}_{k}[n]$, $\overline{r}^{(0)}_{k}[n]$, and $\overline{f}^{(0)}_{k}[n]$ for all $k$ and $n$;
\State Set $i=1$;
\State Given \{$\mathbf{q}^{(i-1)}_{\mathrm{u}}[n],\overline{t}^{(i-1)}_{k}[n],\overline{e}^{(i-1)}_{k}[n]$\} solve problem (19) to obtain \{$P^{(i)}_{k}[n],\mathbf{X}^{(i)}[n],t^{(i)}_{k}[n],e^{(i)}_{k}[n]$\} for all $k$ and $n$;
\State Update \{$\overline{t}^{(i-1)}_{k}[n],\overline{e}^{(i-1)}_{k}[n]$\} $\leftarrow$ \{$t^{(i)}_{k}[n],e^{(i)}_{k}[n]$\} for all $k$ and $n$;
\State \textbf{Repeat} step-3 to step-4 until convergence of the objective value of (19), and return the updated $P^{(i)}_{k}[n]$, $\mathbf{X}^{(i)}[n]$ for all $k$ and $n$;
\State Given \{$P^{(i)}_{k}[n],\mathbf{X}^{(i)}[n],\overline{r}^{(i-1)}_{k}[n]$, $\overline{f}^{(i-1)}_{k}[n]$\}, solve problem (24) to obtain \{$\mathbf{q}^{(i)}_{\mathrm{u}}[n],r^{(i)}_{k}[n],f^{(i)}_{k}[n]$\} for all $k$ and $n$;
\State Update \{$\overline{r}^{(i-1)}_{k}[n],\overline{f}^{(i-1)}_{k}[n]$\} $\leftarrow$ \{${r}^{(i)}_{k}[n],{f}^{(i)}_{k}[n]$\} for all $k$ and $n$;
\State \textbf{Repeat} step-6 to step-7 until convergence of the objective value of (24), and return $\mathbf{q}^{(i)}_{\mathrm{u}}[n]$ for all $n$;
\State Calculate $\overline{t}^{(i)}_{k}[n],\overline{e}^{(i)}_{k}[n],\overline{r}^{(i)}_{k}[n],\overline{f}^{(i)}_{k}[n]$ based on the returned $P^{(i)}_{k}[n],\mathbf{X}^{(i)}[n],\mathbf{q}^{(i)}_{\mathrm{u}}[n]$ for all $k$ and $n$, and let $i=i+1$;
\State \textbf{Repeat} step-3 to step-9 until convergence of the objective value of (14);
\State \textbf{Return:} $P^{*}_{k}[n]$, $\mathbf{X}^{*}[n]$, $\mathbf{q}^{*}_{\mathrm{u}}[n]$ for all $k$ and $n$.
\end{algorithmic}
\end{algorithm}\\
It is worth noting that, to guarantee the feasibility conditions, step-9 in \textit{Algorithm 1} can be done by replacing `$\geq$' with `$=$' in (15b) for calculating $\overline{t}^{(i)}_{k}[n]$ and replacing `$\leq$' with `$=$' in (17) for calculating $\overline{e}^{(i)}_{k}[n]$. In addition, $\overline{r}^{(i)}_{k}[n]$ and $\overline{f}^{(i)}_{k}[n]$ can be calculated by following the same way to calculate $\overline{r}^{(0)}_{k}[n]$ and $\overline{f}^{(0)}_{k}[n]$. 

The computational complexity of \textit{Algorithm 1} can be summarised as: Problem (19) is a semidefinite optimization problem (SDP), problem (24) is a second-order cone optimization problem (SOCP), and most of toolboxes handle SDP/SOCP using an interior-point method \cite{Boyd2004}. In this case, the worst-case complexity for solving problem (19) is $\mathcal{O}(\mathcal{I}_{1}\log(1/\epsilon)\max\{3KN,2MN\}^{4})$ and for solving problem (24) is $\mathcal{O}(\mathcal{I}_{2}\log(1/\epsilon)(2KN)^{3})$, where $\mathcal{I}_{1}$ and  $\mathcal{I}_{2}$ denote the number of iterations for the SCA in solving (19) and (24), respectively; $\epsilon$ denotes the algorithm stopping criteria. Thus, considering a similar value for both $\mathcal{I}_{1}$ and $\mathcal{I}_{2}$, the worst-case complexity of the proposed \textit{Algorithm 1} can be approximated as $\mathcal{O}(\mathcal{I}_{3}\mathcal{I}_{1}\log(1/\epsilon)\max\{3KN,2MN\}^{4})$, where $\mathcal{I}_{3}$ is the number of iterations for one complete loops search.



\section{Numerical Results}
In this section, we provide several experiments to examine the proposed solution for the UAV energy minimization problem. Suppose that there are $K=4$ IoT devices distributed within a geographic area of size $2\times2~\mathrm{m}^2$ following an independently Poisson point process \cite{Hou2018}. The MIMO FD UAV is equipped with $M=4$ antennas flying at the altitude $L=2~\mathrm{m}$, starting from the point $\mathbf{q}_{\mathrm{ui}}=[-1,-1]$ and ending at the point $\mathbf{q}_{\mathrm{uf}}=[1,-1]$. The maximum UAV flying speed is $V_{\mathrm{max}}=10~\mathrm{m/s}$ and its transmit power is limited to $P_{\mathrm{max}}=10~\mathrm{W}$. We set the total flying time $T=60~\mathrm{s}$ and it is divided by $N=8$ time slots. The communication bandwidth is $B=10~\mathrm{Hz}$ and the number of uploaded data per IoT device is $R_{k}=256~\mathrm{bits}, \forall k$. The hardware transceiver distortion coefficients of the UAV are set to $\kappa=0.005$ and $\beta=0.01$. The IoT devices energy harvesting and the UAV speed efficiencies are set to $\eta=\tau=0.6$. 

Furthermore, two trajectory benchmark schemes are considered: For benchmark 1, the UAV flies to the centre point of the geographic area, i.e., [0,0], and hovers there for all $N$ time slots for data collection; For benchmark 2, the UAV first flies to the point [-0.7778,0], and then along x-axis flies straight to the point [0.7778,0], and finally it flies back to the ending point. In this case, the UAV collects data at $N$ even points between [-0.7778,0] and [0.7778,0]. It is worth noting that the above configurations are purely for demonstrating our proposed idea. For a practical experimental set up, we only need to linearly scale the above mentioned parameters. 
 
\begin{figure}[t] 
\begin{center}
\epsfig{figure=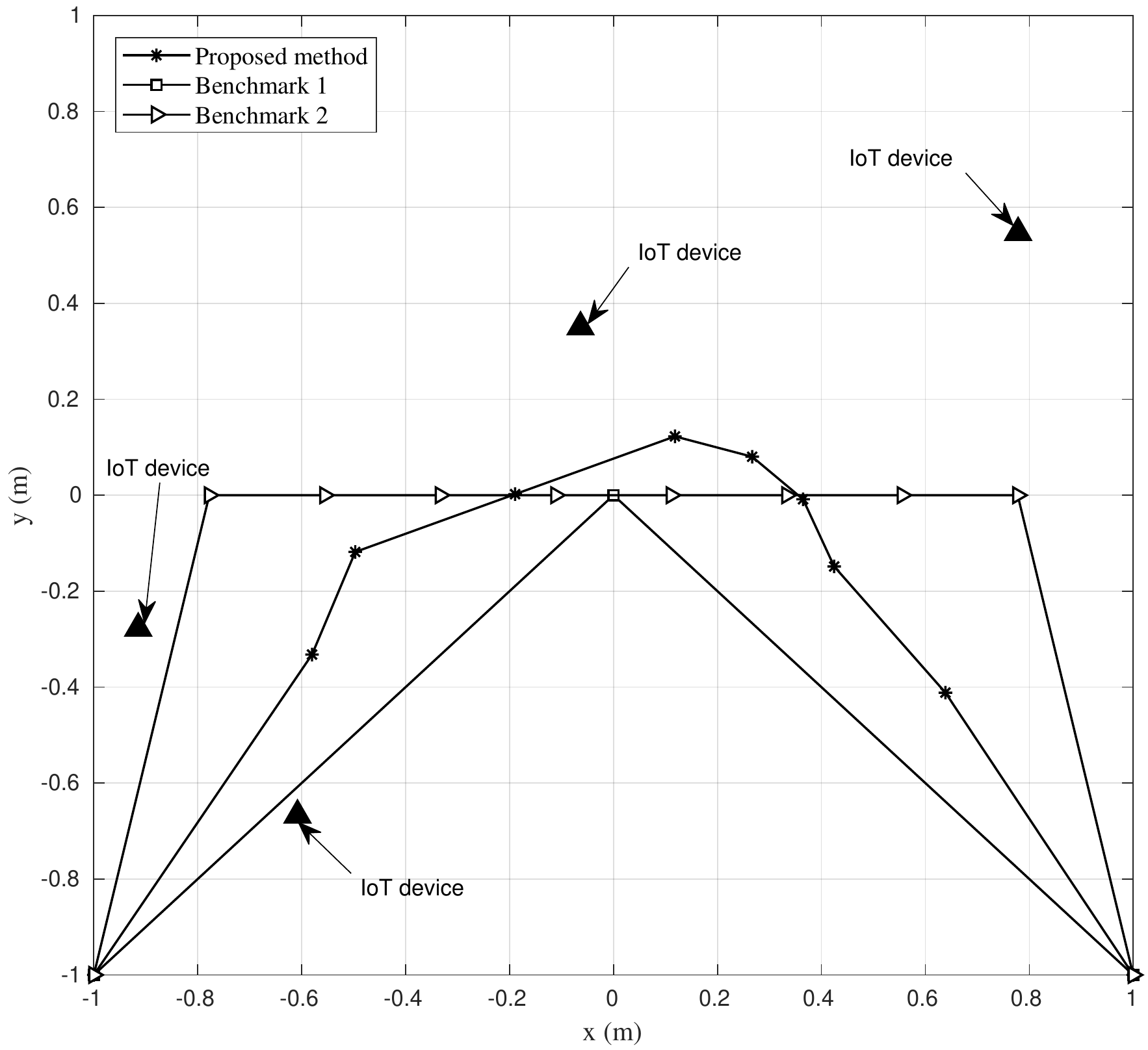,scale=0.43,angle=0}
\end{center}
\caption{The optimized UAV trajectory projected on the horizontal plane, where Rician fading $K$-factor is 0.1 for IoT devices channels and 1 for SI channels, and $\sigma^{2}_{0}=0.01.$}\label{F1}
\end{figure}

Fig.~\ref{F1} gives our optimized UAV trajectory projected on the horizontal plane comparing with two benchmark schemes.
As shown in the figure, the optimized UAV trajectory for $N$ time slots data collection is almost between the two benchmark trajectories. In detail, the UAV first hovers between the left two IoT devices for two time slots data collection. Then, it flies towards the IoT device near the coordinate point [-0.05,0.3] for another two time slots data collection. Next, it spends another three time slots hovering towards the last IoT device direction in the right-up corner of the geographic area for data collection. In this case, since the UAV must fly back to the ending point, it cannot hover quite close to the last IoT device for its data collection.  

\begin{figure}[t] 
\begin{center}
\epsfig{figure=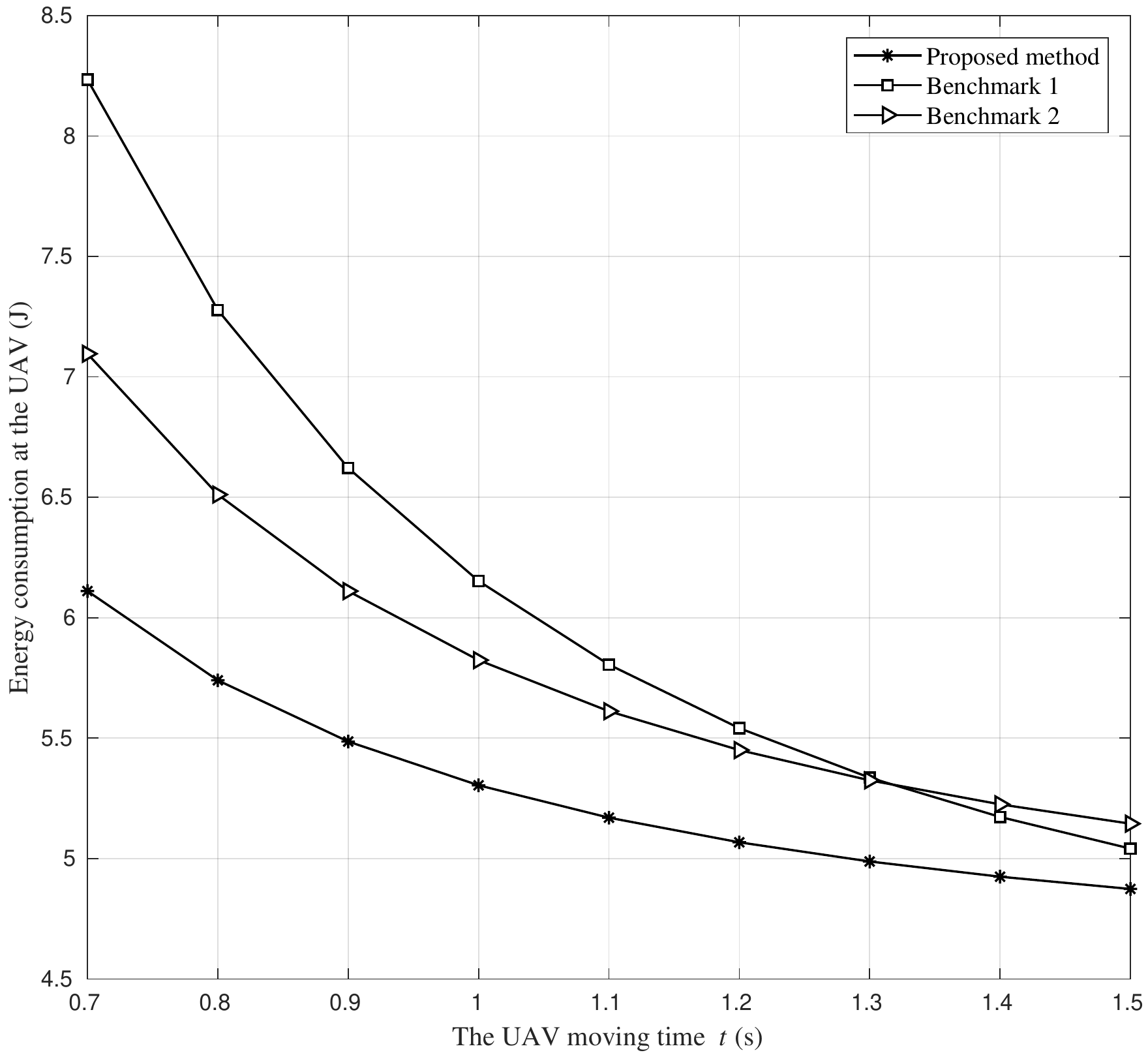,scale=0.43,angle=0}
\end{center}
\caption{Energy consumption at the UAV versus the UAV moving time $t$, where Rician fading $K$-factor is 0.1 for IoT devices channels and 1 for SI channels, and $\sigma^{2}_{0}=0.01$.}\label{F2}
\end{figure}

Fig.~\ref{F2} shows energy consumption of our optimized UAV trajectory in comparison with the two benchmark schemes for different UAV moving time $t$.
As shown in the figure, by comparing with the two benchmark schemes, the proposed method that leads to the optimized UAV trajectory consumes the smallest amount of energy of the UAV. This demonstrates that our proposed method jointly optimizes the data collection, WPT, and the UAV trajectory in terms of the energy minimization of the UAV. In addition, accompanied with the UAV moving time increasing, the energy consumptions at the UAV for all three schemes are decreasing. This is because the speed of the UAV is decreasing with a fixed distance between two hovering locations, i.e., see \eqref{eq12}. In other words, the UAV propulsion energy is decreasing accompanied with the UAV moving time $t$ increasing. Moreover, benchmark 2 shows less energy consumption than benchmark 1 when the UAV moving time $t$ is smaller than 1.32 s. This is because the distance between the starting point and the hovering point for benchmark 1 is much larger than the ones for benchmark 2, where, with a fixed $t$, the former one leads to a large UAV propulsion energy consumption. On the other hand, when $t$ is getting large, the energy consumption in terms of WPT will domain the performance. In this case, benchmark 1 hovering in the centre point of the geographic area will provide more energy-efficient data collection and WPT.   

\begin{figure}[t] 
\begin{center}
\epsfig{figure=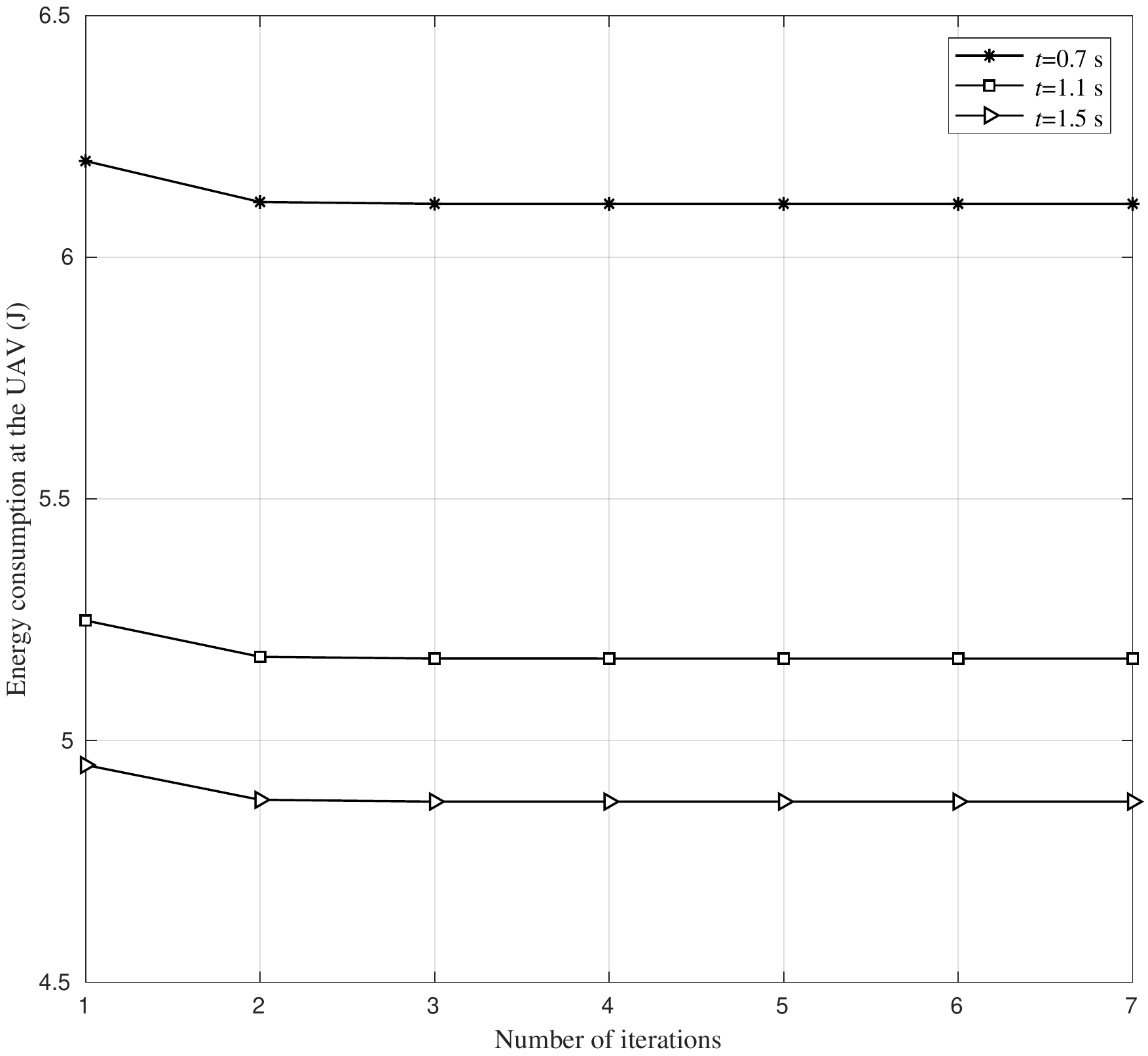,scale=0.43,angle=0}
\end{center}
\caption{Energy consumption at the UAV versus the number of complete loops iterations with different UAV moving time $t$.}\label{F3}
\end{figure}

Fig.~\ref{F3} then demonstrates the convergence property of the proposed iterative search algorithm.
As shown in the figure, for different UAV moving time $t$, around 4 complete loops iterations (i.e., $\mathcal{I}_{3}$) will lead to the optimal solution, and combined with the analysis in Sec. IV-C, the proposed algorithm indeed leads to relatively low computational complexity.



\section{Conclusion}
In this paper, we have proposed an energy-efficient data collection and WPT framework for IoT applications via a MIMO FD UAV. By formulating and solving the energy minimization problem in consideration of the UAV hardware impairments, we optimized the UAV trajectory, the covariance matrix of UAV energy signal for WPT, and the transmit power of each IoT devices. It has been shown in numerical results that, our proposed algorithm outperforms the two presented benchmark schemes in terms of UAV energy efficiency with low computational complexity and fast convergence.



\section*{Acknowledgement}
This work was supported by the Engineering and Physical Science Research Council (EPSRC) through the Scalable Full Duplex Dense Wireless Networks (SENSE) grant EP/P003486/1.


\bibliography{mybib}
\bibliographystyle{IEEEtran}
\end{document}